%% file: quantum-template.tex
\documentclass[a4paper,twocolumn,11pt, unpublished]{quantumarticle}
\pdfoutput=1
\usepackage[utf8]{inputenc}
\usepackage[english]{babel}
\usepackage[T1]{fontenc}
\usepackage{hyperref}
\usepackage{xcolor}
\usepackage{tikz}
\usepackage{lipsum}

\usepackage[utf8]{inputenc}
\usepackage[english]{babel}
\usepackage[T1]{fontenc}
\usepackage{amsmath, amssymb}
\usepackage[numbers,sort&compress]{natbib}
\usepackage{graphicx}
\usepackage{pbox}
\usepackage{parskip}

\graphicspath{{./Images/}} 
\usepackage{caption}         
\usepackage{subcaption}

\begin{document}

\title{Restricted Global Optimization for QAOA}

\author{Peter Gleißner}
\affiliation{Fraunhofer Institute for Integrated Systems and Device Technology IISB, Erlangen, Germany}
\email{peter.gleissner@tu-ilmenau.de}
\author{Georg Kruse}
\affiliation{Fraunhofer Institute for Integrated Systems and Device Technology IISB, Erlangen, Germany}
\email{georg.kruse@iisb.fraunhofer.de}
\author{Andreas Roßkopf}
\affiliation{Fraunhofer Institute for Integrated Systems and Device Technology IISB, Erlangen, Germany}
\email{andreas.rosskopf@iisb.fraunhofer.de}
\maketitle
\maketitle

\begin{abstract}
The Quantum Approximate Optimization Algorithm (QAOA) has emerged as a promising variational quantum algorithm for addressing NP-hard combinatorial optimization problems. However, a significant limitation lies in optimizing its classical parameters, which is in itself an NP-hard problem. To circumvent this obstacle, initialization heuristics, enhanced problem encodings and beneficial problem scalings have been proposed. While such strategies further improve QAOA's performance, their remaining problem is the sole utilization of local optimizers. We show that local optimization methods are inherently inadequate within the complex cost landscape of QAOA. Instead, global optimization techniques greatly improve QAOA's performance across diverse problem instances. While global optimization generally requires high numbers of function evaluations, we demonstrate how restricted global optimizers still show better performance without requiring an exceeding amount of function evaluations.
\end{abstract}

\input{introduction}
\input{related_works}

\input{cost_landscape}
\input{global_optimization}
\input{experiments}

\input{conclusion}

\section*{Acknowledgement}
This project is supported by the Federal Ministry for Economic Affairs and Climate Action on the basis of a decision by the German Bundestag through the project Quantum-enabling Services and Tools for Industrial Applications (QuaST).

\bibliographystyle{quantum}
\bibliography{bibliography}

\onecolumn\newpage

\appendix
\input{appen_opt}

\end{document}

%% file: introduction.tex
\section{Introduction}

Quantum algorithms like Shor's algorithm \cite{shor} and the Harrow-Hassidim-Lloyd (HHL) algorithm \cite{hhl} show exponential advantages over their classical counter parts. Since such quantum algorithms demand fault-tolerant quantum computers, variational quantum algorithms have emerged for currently available Noisy Intermediate-Scale Quantum (NISQ) devices, which combine parameterized quantum circuits and classical optimizers. One of these variational quantum algorithms is the Quantum Approximate Optimization Algorithm (QAOA), which is considered a promising candidate to show quantum advantage for NP-hard combinatorial optimization problems. While extrapolations suggest a quantum speed-up for problem instances that require several hundreds of qubits \cite{speed-up}, the main limitation lies in the exponential cost of optimizing the classical parameters of this quantum algorithm, which has been shown to be a NP-hard problem itself \cite{np}. The cost landscape of QAOA exhibits wide plane areas and large numbers of local minima \cite{Ge.19.01.2022}, which are especially disadvantageous for local optimizers. To overcome this issue, heuristic strategies for the classical optimization as well as favorable problem formulations and encodings have been studied \cite{Brandhofer.2023, Zhou.2020, Nakanishi.2020, Sack.2021, Streif.23.08.2019}. While such strategies enhance the performance of QAOA, further improvements are crucial in order to apply QAOA to real-world optimization problems like the Unit Commitment (UC) problem \cite{Koretsky.17.10.202122.10.2021}, the Traveling Salesperson (TSP) problem \cite{Palackal.19.04.2023} or the Factory Layout (FL) problem \cite{Klar.2022}. 

In classical optimization, it is well known that optimization algorithms, which are solely based on local information, are generally not adequate for solving NP-hard problems \cite{global_op1}. Instead, global optimization algorithms which combine  multi-point global search and problem oriented local search  have been well established for such problems  \cite{global_op2, global_op3}.

In this work we show why current approaches with local optimizers show poor performance, even for simple problem instances. While a current line of research aims to improve the overall structure of the cost landscape \cite{Brandhofer.2023, Zhou.2020}, we argue that such improvements cannot resolve the drawbacks caused by the use of local optimizers. We therefore introduce global optimizers and show that they greatly improve the performance of QAOA on various problem instances and sizes, at the cost of a higher number of function evaluations.

Our work is structured as follows: In Section \ref{chap_rel_works} we motivate the need for better optimization techniques and discuss current approaches. In Section \ref{preliminaries} we briefly discuss QAOA and the problem formulations used in this work. In Section \ref{chap_cl_cost_topology}, we show how the cost landscape of general problem formulations undergoes a transformation based on both the selected encoding parameters as well as the general parameter choices of QAOA. In Section \ref{chap_global_optimization} we motivate the use of global optimizers on the basis of a simple example and benchmark their performance on three use cases in Section \ref{chap_exp} and \ref{chap_results}. Finally, we conclude our findings in Section \ref{chap_conclusion}.

%% file: related_works.tex
\section{Related Works}\label{chap_rel_works}

To overcome the difficulty of optimizing the classical parameters of QAOA, various optimization strategies as well as favorable problem encodings and formulations have been proposed. One line of research focuses on improving the structure of QAOA's cost landscape in order to reduce its difficulty for local optimizers. Brandhofer et al. show how different factors like scaling parameters, encoding strategies and mixer Hamiltonians can influence the shape of the cost landscape and therefore influence optimization performance \cite{Brandhofer.2023}. Albeit many enhancements have been proposed, this line of research does not consider the underlying problem of optimizing QAOA's classical parameters with local optimizers.

Another line of research proposes heuristics for parameter initialization in order to minimize the effort of classical optimization. Nakanishi et al. combine such a heuristic with a layer-wise training approach \cite{Nakanishi.2020}, where each layer of QAOA is optimized after the other, leading to improved results at the cost of higher number of optimization steps. Zhou et al. also propose a heuristic for a layer wise initialization \cite{Zhou.2020}, while Sack et al. propose adapting concepts from Quantum Annealing to the initialization of classical parameters \cite{Sack.2021}.
A similar approach is the idea of parameter transfer, where the classical parameters of QAOA are first optimized for a simpler problem instance and the resulting parameters are then used for the initialization of harder problem instances \cite{Streif.23.08.2019, Shaydulin.2023, Harrigan.2021}. 

Instead of improving optimization results of QAOA by enhanced encodings or heuristics, various works have tried to replace the local optimizers by other optimization approaches such as reinforcement learning  \cite{Khairy.2020}. Closest to our claims are the works of \cite{genetic} and  \cite{Rad.04.03.2022}. Acampora et al. demonstrate how the use of evolutionary strategies enhances the performance of QAOA on the Max-Cut problem, and Rad et al. show how the optimization of classical parameters of QAOA can be separated into two distinct phases: a first global phase where a bayesian estimate is used to initialize the parameters of a second local phase. Both works, which can be considered to use global optimization approaches, show enhanced performance over local optimizers.

Besides these approaches, many concepts like warm-starting \cite{warm-start} have been introduced, which improve the solution quality of QAOA without addressing the difficulty of optimizing the classical parameters with local optimizers.

%% file: cost_landscape.tex
\section{Preliminaries}\label{preliminaries}

The cost landscape of hybrid quantum-classical optimization algorithms like QAOA  has been subject to detailed analyses. The specific shape changes with the problem at hand and its encoding into a Hamiltonian. To start from a common basis, Section \ref{chap_cl_QAOA_impl} introduces the general structure of QAOA and Section \ref{chap_cl_problem_form} the general form of the problem formulations used in this work.

\subsection{QAOA}\label{chap_cl_QAOA_impl}

The aim of the optimization algorithm is the preparation of a state $\vert\psi(\beta,\gamma)\rangle$ that encodes the solution to the given problem, where $\beta$ and $\gamma$ are parameter vectors that are optimized by a classical optimizer. The output state is prepared from an equal superposition by an Ansatz that consists of layers. Each of them depends on the value of $\beta_i$ and $\gamma_i$, where $i$ denotes the index of the layer. The layers are repeated $p$ times, with $p$ being small for implementations on NISQ devices. Each layer $\hat{H}_{layer}$ consist of two parts: The cost and the mixing Hamiltonians $\hat{H}_C$ and $\hat{H}_M$ respectively. 

\begin{equation}
\hat{H}_{layer}=e^{-i\beta_i\hat{H}_M}e^{-i\gamma_i\hat{H}_C}
\end{equation}
Here, $\beta_i$ parameterizes the mixer term, while $\gamma_i$ parameterizes the cost term.

The complete circuit can be written as
\begin{multline}
\vert\psi(\beta,\gamma)\rangle = e^{-i\beta_p\hat{H}_M}e^{-i\gamma_p\hat{H}_C}\\ ... e^{-i\beta_1\hat{H}_M}e^{-i\gamma_1\hat{H}_C}\vert+\rangle^{\otimes n}.
\end{multline}

For all cases considered in this work, the mixer Hamiltonian is defined simply as
\begin{equation}
\hat{H}_M=\sum_i \hat{\sigma}_i^x .
\end{equation}

There are several other options to chose the mixer Hamiltonian to improve performance \cite{Brandhofer.2023}, but in this work we want to focus on the general drawbacks of local optimizers and therefore only consider this simplified case.

The cost Hamiltonian on the other hand encodes the given problem and can be constructed from a quadratic unconstrained binary optimization (QUBO) formulation via the Ising-Model. The cost Hamiltonian will be constructed as

\begin{equation}
\hat{H}_C = \sum_{j<i}J_{ij}\hat{\sigma}_i^z\hat{\sigma}_j^z+\sum_{i} h_i \hat{\sigma}_i^z ,
\end{equation}

with $J_{ij}$ being the factors of the quadratic terms of the $i$-th and $j$-th element and $h_i$ the factors of the linear terms of the Ising-Model. The so constructed circuit generates states depending on the parameters $\beta$ and $\gamma$. The quality of a given state depends on the expectation value of the cost function denoted as $C(\beta,\gamma)$. This cost function is minimized by the classical optimizer.

\subsection{Problem Formulation}\label{chap_cl_problem_form}

The problem formulations of interest consist of an objective function and a set of constrains for a binary (or integer) optimization problem. In order to encode this problem formulation into the cost Hamiltonian for QAOA, it needs to be converted into a QUBO formulation. A general guide on doing this can be found in \cite{Glover.13.11.2018}. In this work the cost function of a given QUBO will have the following general form:

\begin{multline}\label{eq_gen_QUBO}
c(\{x_i\}) = s(H_{cost}(\{x_i\})+ \\
\sum_{j=1}^{n_{pen}}P_jH_{pen,j}(\{x_i\})) .
\end{multline}

Here $x_i$ refers to the binary variables of the optimization problem. $H_{cost}$ and $H_{pen,j}$ are the functions encapsulating the cost and the $n_{pen}$ penalties. The factor $s$, as described in \cite{Shaydulin.2023}, is a scaling factor that is chosen, such that it ensures numerical stability in the optimization process (ref. Section \ref{chap_cl_s}). Lastly, $P_j$ scales the respective penalty term to change the influence it has on the cost landscape as shown in \cite{Palackal.19.04.2023, Hodson.13.11.2019, Brandhofer.2023} (ref. Section \ref{chap_cl_P}). A detailed description of the problem formulations of the UC, TSP and FL problem can be found in Appendix \ref{appendix_problems}.

\subsection{Performance Metrics}

The performance of an optimization algorithm depends on two main factors: The quality of the solution found and the calculation cost of this solution. The latter can be easily quantified in our setting as the number of cost function evaluations (which corresponds to the number of circuit executions). Therefore, the number of function evaluations will be the measure of how costly it is to obtain a given solution. The quality of the solution is measured by the expectation value of the cost function divided by the cost of the minimal solution, so the normalized cost $\frac{C(\beta,\gamma)}{C_{min}}$ \cite{Harrigan.2021}. This metric has the benefit, that it is not dependent on the absolute value of the cost function. Due to the dropping of the constant offset in the construction of the cost function the minimal cost will always be negative, so the optimal value is 1. Given these performance metrics, a given optimizer should maximize the normalized cost while minimizing the number of function evaluations.

\begin{figure*}[ht!]
    \centering 
\begin{subfigure}{0.49\textwidth}
  \includegraphics[width=\linewidth,trim= 12 37 32 40,clip]{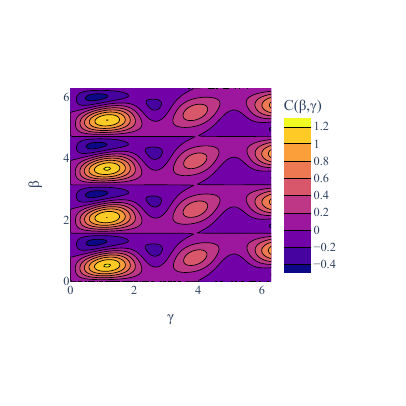}
  \caption{Baseline $s=0.0002$, $P=0.2$, $L=300$}
  \label{fig_pd_base}
\end{subfigure}\hfil 
\begin{subfigure}{0.49\textwidth}
  \includegraphics[width=\linewidth,trim= 12 37 32 40,clip]{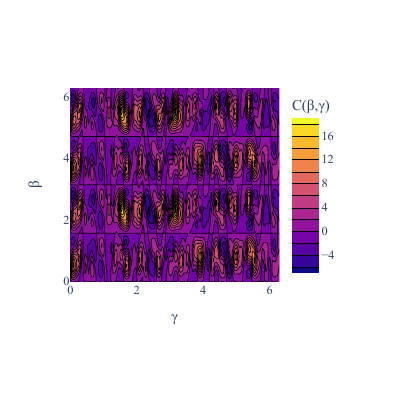}
  \caption{10-times higher $P$}
  \label{fig_pd_P}
\end{subfigure}

\medskip
\begin{subfigure}{0.49\textwidth}
  \includegraphics[width=\linewidth,trim= 12 37 32 40,clip]{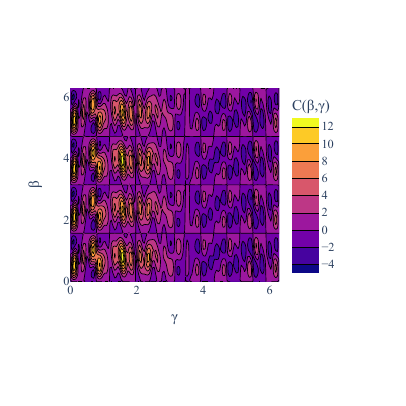}
  \caption{10-times higher $s$}
  \label{fig_pd_s}
\end{subfigure}\hfil 
\begin{subfigure}{0.49\textwidth}
  \includegraphics[width=\linewidth,trim= 12 37 32 40,clip]{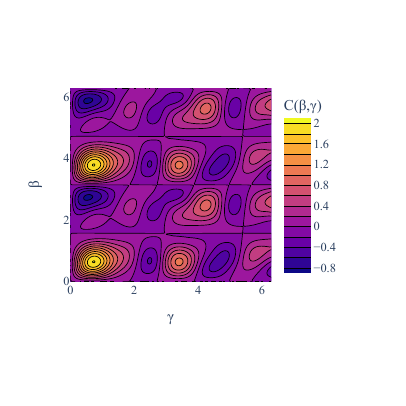}
  \caption{$L=200$}
  \label{fig_pd_L}
\end{subfigure}
\caption{Influence of problem dependent parameters: All contour plots depict cost landscapes created by a four Unit UC problem. Subfigure \ref{fig_pd_base} serves as a \textit{best case} baseline cost landscape with parameters $s=0.0002$, $P=0.2$ and $L=300$ (chosen according to Section \ref{chap_cl_cost_topology}). Subfigures \ref{fig_pd_P} and \ref{fig_pd_s} show the introduction of additional minima with a changed values of $P=2$ and $s=0.002$ respectively. Subfigure \ref{fig_pd_L} shows an additional cost landscape created by a related problem with $L=200$.}
\label{fig_problem_dependent}
\end{figure*}

\section{Cost Landscape}\label{chap_cl_cost_topology}

The shape of the cost landscape for a given problem depends on a set of parameters that are influenced both by the specific problem at hand and the characteristics of the algorithm being used. These parameters greatly influence the performance of classical optimization methods and require a thorough examination of their impact on the underlying cost landscape. As we show in Section \ref{chap_global_optimization}, even for well designed QAOA cost landscapes, optimization (especially for local optimizers) remains difficult. To obtain such well designed QAOA cost landscapes, the effects of various parameters will be studied in the following. 

Problem dependent parameters like the power demand $\mathnormal{L}$ of the UC problem or the adjacency matrix $\mathnormal{D}$ of the TSP and FL problem are directly tied to the problem. Other parameters, like the scaling factor $\mathnormal{s}$ or pentalty factors $\mathnormal{P_i}$, have to be chosen such that stable and successful optimization is achieved. On the other hand, algorithmic dependent parameters are (partly) independent of the problem instance and influence the characteristics of the QAOA cost landscape as well. In this work only qubit number $\mathnormal{n}$ and layer count $\mathnormal{p}$ are investigated, dropping the additional degrees of freedom stemming from the choice of mixer Hamiltonian and enhanced encoding strategies.

The visualization of the cost landscapes for a given problem formulation for QAOA with one layer is straight forward: The parameters  $\beta$ and $\gamma$ are used as $x$ and $y$ axis respectively and the expectation value of the cost function $C(\beta, \gamma)$ is used as $z$ axis (ref. Figure \ref{fig_problem_dependent}). In order to plot higher dimensional cost landscapes we follow the ideas of \cite{Li.28.12.2017} where two random vectors ($\theta_1$ and $\theta_2$) serve as axes needed to visualize parts of the cost landscape.  

A characteristic example of a well designed cost landscape of QAOA with a single layer for a four Unit UC problem instance is depicted in Figure \ref{fig_pd_base}: Due to the simple construction of the mixer Hamiltonian as described in \cite{Sack.2021}, a clear periodicity in the direction of $\beta$ can be seen. This is not the case in the direction of $\gamma$, where (depending on the Hamiltonian) many different frequency components will overlay \cite{Stechy.23.05.2023}. Generally, the detection of the combined period for arbitrary Hamiltonians is unlikely. Again this can be seen in Figure \ref{fig_pd_base}. Since the periodicity of $\gamma$ for the Hamiltionians considered in this work is unknown, strategies of using parameter initializations similar to an annealing schedule, as proposed in \cite{Sack.2021}, cannot be applied. 

The similarity of the position of local minima in the cost landscapes depicted in Figure \ref{fig_problem_dependent} leads to a path of investigation where one tries to apply the knowledge of one found minimum in one problem instance to other problem instances. This includes the concepts from \cite{Streif.23.08.2019,Shaydulin.2023,Jing.2023} where this strategy is successfully demonstrated. These concepts have the caveat, that they need prior knowledge of solutions found for similar problems, so the initial problem of finding a primary good solutions remains the same, motivating the use of global optimizers.

\begin{figure*}[ht!]
    \centering 
    
\begin{subfigure}{0.49\textwidth}
  \includegraphics[width=\linewidth,trim= 12 37 32 40,clip]{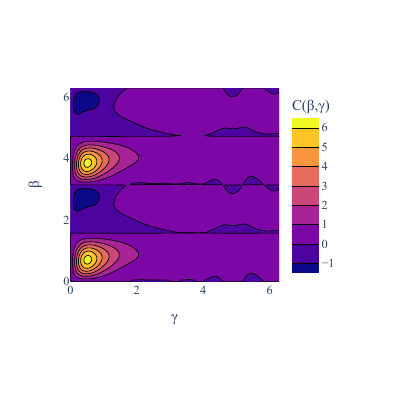}
  \caption{1 layer, 8 qubits}
  \label{fig_ad_8q}
\end{subfigure}\hfil 
\begin{subfigure}{0.49\textwidth}
  \includegraphics[width=\linewidth,trim= 12 37 32 40,clip]{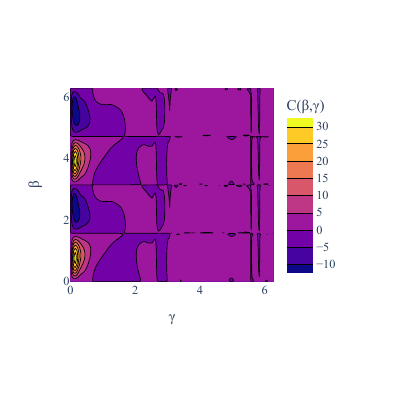}
  \caption{1 layer, 12 qubits}
  \label{fig_ad_12q}
\end{subfigure}

\medskip
\begin{subfigure}{0.49\textwidth}
  \includegraphics[width=\linewidth,trim= 12 37 32 40,clip]{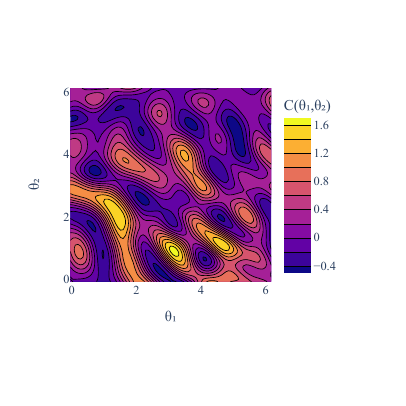}
  \caption{3 layers, 4 qubits}
  \label{fig_ad_3l}
\end{subfigure}\hfil 
\begin{subfigure}{0.49\textwidth}
  \includegraphics[width=\linewidth,trim= 12 37 32 40,clip]{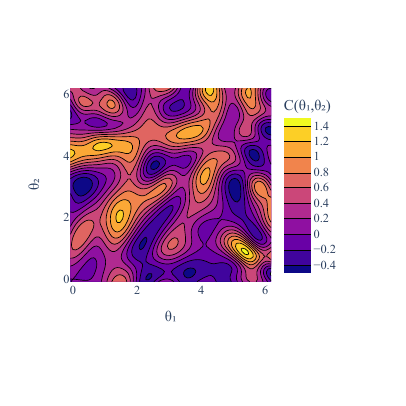}
  \caption{5 layers, 4 qubits}
  \label{fig_ad_5l}
\end{subfigure}

\caption{Influence of algorithmic dependent parameters: All contour plots depict cost landscapes created by the UC problem at $s=0.0002$, $P=0.2$ and $L=300$ (same as baseline in Figure \ref{fig_pd_base}). Subfigures  \ref{fig_ad_8q} and \ref{fig_ad_12q} show the effect more qubits have on the landscape. Subfigures \ref{fig_ad_3l} and  \ref{fig_ad_5l} show the impact of an increase in layers, with the axis being random vectors in the parameter space.}
\label{fig_algorithmic_dependent}
\end{figure*}

\subsection{Effect of Scaling Factor $s$}\label{chap_cl_s} 

The magnitude of $s$ (ref. Equation \ref{eq_gen_QUBO}) has a direct effect on the scaling of the coefficients in the cost Hamiltonian. As described by \cite{Shaydulin.2023}, this stretches or compresses the cost landscape with respect to $\gamma$, when $s$ is smaller or greater than 1 respectively. The comparison between Figure \ref{fig_pd_base} and \ref{fig_pd_s} illustrates this phenomenon. The change in $s$ by a factor of $10$ compresses the landscape spanning from $0$ to $2\pi$ down to $0$ to around $0.4$. Additionally, the magnitude of $C(\beta,\gamma)$ also changes, as the calculation of the cost function is also dependent on $s$. The resulting states and their corresponding solutions are not changed, so the minima remain the same, solely shifted by this parameter.

As the effectiveness of standard optimization algorithms often deteriorates if the variables do not have the same order of magnitude, $s$ should be chosen such that all variables fullfill this requirement. In figure \ref{fig_pd_s} a further increase of $s$ would result in a cost landscape, where small changes in $\gamma$ would have an even greater effect, making convergence difficult. In this work, the rescaling heuristic suggested in \cite{Shaydulin.2023} is used to enhance the cost landscape, where $s$ is chosen such that the absolute mean weight of the coefficients is scaled to 1. This ensures that at least one minimum lies in the range $\gamma \in [0, 2\pi]$, but also results in a trade-off between excluding better minima from the search space on one side and a smaller, but limited parameter space on the other side.

\subsection{Effect of Penalty Factors $P_j$}\label{chap_cl_P}

The penalty factors affects the number and size of the local minima: With rising  values of $P_j$, more local minima are placed in the same segment of the parameter space. This is due to the increased value a invalid solution adds to the expectation value of the solution, as states representing wrong solutions become less desirable. This effect is depicted in Figures \ref{fig_pd_base} and \ref{fig_pd_P}, where the increased $P$ results in a rougher surface. When choosing the value of $P_j$ there is again a trade-off: If $P_j$ is too low,  invalid solutions will become optimal (in terms of $H_{cost}$). If $P$ is too large, the cost landscape becomes difficult to optimize in. Additionally, the relative difference between valid and optimal solutions decreases in comparison to the over all values of the cost landscape, so valid and optimal solutions become increasingly similar. 

Works such as \cite{Brandhofer.2023, Hodson.13.11.2019} have proposed methods to choose a favorable penalty factor for a given problem. In this work we follow the strategy suggested in \cite{Brandhofer.2023}, where $P_j$ is iteratively increased by small values until the cheapest wrong (invalid) solution becomes at least as expensive as the optimal solution. A short synopsis to this method  can be found in Appendix \ref{append_ah_P}. In real applications this strategy is not feasible as it requires knowledge of the full cost structure of the problem.

\subsection{Effect of the Problem Dependent Variables}\label{chap_cl_L}

In contrast to $s$ and $P_j$, problem dependent variables such as the power demand $L$ in the UC problem can not be chosen freely, as they are part of the problem at hand. The position of the minima of the cost landscape move depending on such values. To illustrate such changes, the power demand $L$ of the UC problem has been varied between Figures \ref{fig_pd_base} and \ref{fig_pd_L}. While the underlying structure of the cost landscape remains the same, the distinct minima change. This also shows the difficulty of applying parameter transfer \cite{Streif.23.08.2019, Shaydulin.2023, Jing.2023} for such problem instances, where some of the minima are greatly shifted or even disappear for varying problem dependent variables.

\subsection{Effect of Qubit Number}

The increase of the number of qubits moves the archetype of the cost landscape from a turbulent shape with many local minima (Figure \ref{fig_pd_base}) to the shape of a barren plateau. Here, local minima occupy a smaller parameter subspace, while wide planes show low variance in the value of the cost function, causing the optimization in this domain to become difficult. This phenomenon can be seen in both Figures \ref{fig_ad_8q} and \ref{fig_ad_12q}, where the amount of minima drastically decreases at the cost of wide planes without any visible structure.

\subsection{Effect of Layer Number}

The effect of the number of layers is the hardest to access with the used method of visual analysis, as a higher number of layers results in higher dimensionality. Therefore, only limited information can be extracted from this comparison. The analysis shows, that the cost landscape exhibits an increased number of local minima,  which is in line with the hypothesis from \cite{Baker.14.02.2022}. An increase of number of free parameters makes the process of optimization increasingly complex. The two examples shown in Figures \ref{fig_ad_3l} and \ref{fig_ad_5l} demonstrate this increased complexity. 

%% file: global_optimization.tex
\section{Global optimization}\label{chap_global_optimization}

Global optimization aims to find the best solution within the entire feasible search space. The search space encompasses all possible solutions of the problem and global optimization attempts to identify the global minimum by considering all variable combinations and constraints. However, global optimizers are often limited to finding the minimum in a designated area, leading to no guarantee of finding the global minimum.

\begin{figure*}[ht!]
    \centering 
\begin{subfigure}{0.5\textwidth}
  \includegraphics[width=\linewidth,trim= 0 0 20 10,clip]{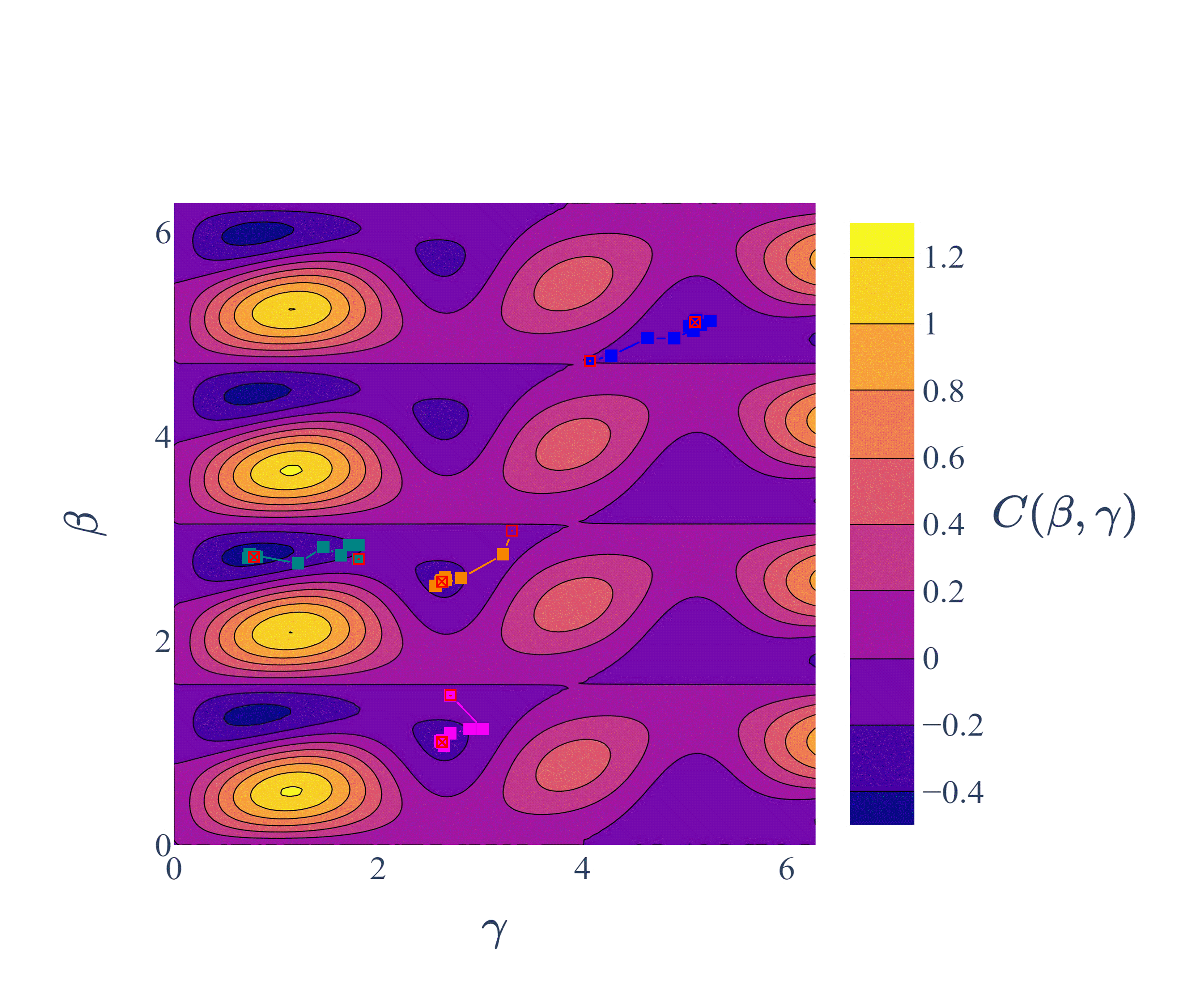}
  \caption{Local optimization}
  \label{fig_cls_NM_progress}
\end{subfigure}\hfil 
\begin{subfigure}{0.5\textwidth}
  \includegraphics[width=\linewidth,trim= 0 0 20 10,clip]{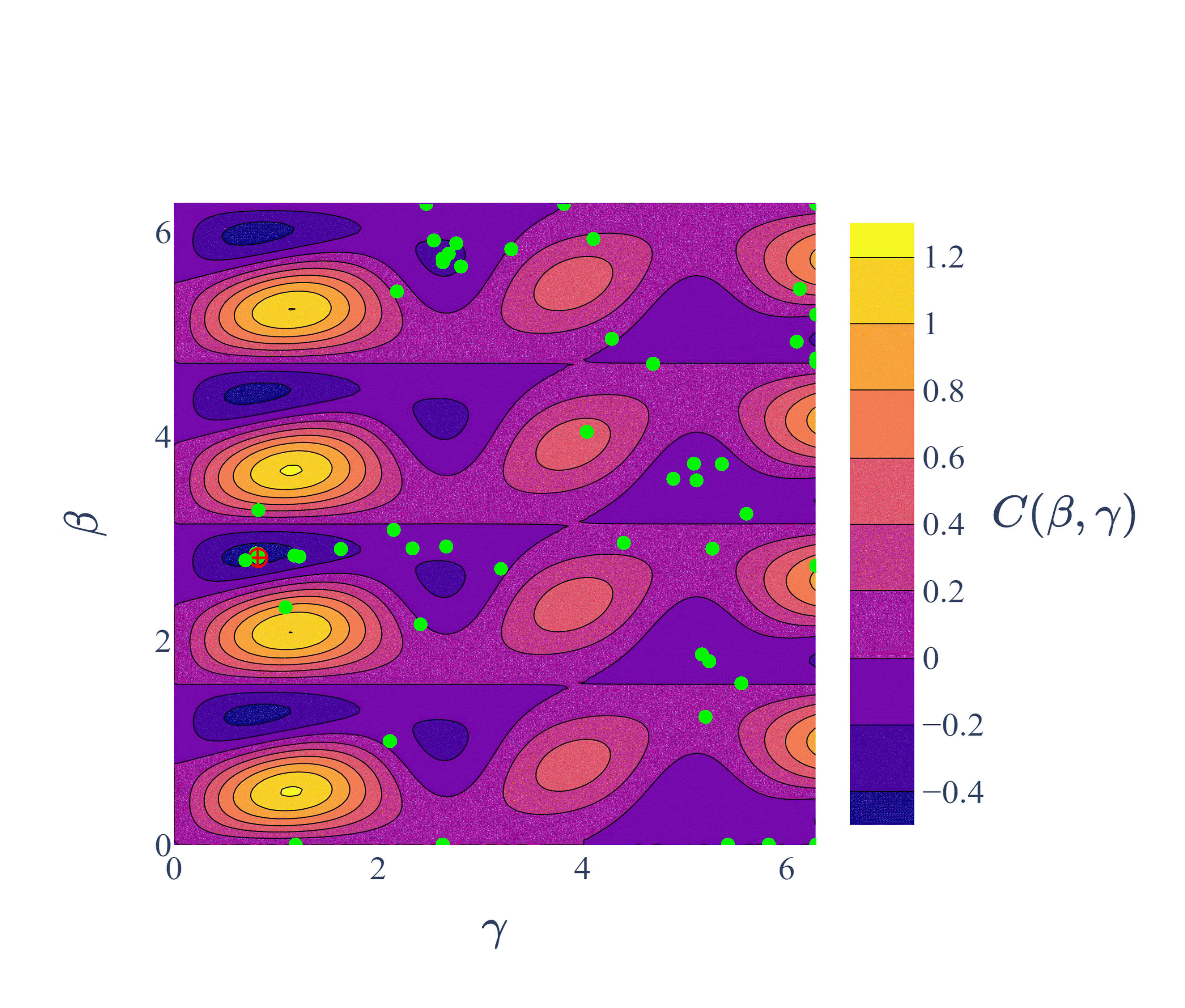}
  \caption{Global optimization}
  \label{fig_cls_FS_progress}
\end{subfigure}\hfil 
\caption{Both surface plots show the cost landscape of the one layer, four unit UC problem baseline with $L=300$, $s=0.0002$ and $P=0.2$ (as in Figure \ref{fig_pd_base}). Subfigure \ref{fig_cls_NM_progress} shows four optimization runs of the randomly initialized local optimizer NM, with the starting and final points indicated by a red marker. Depending on the initialization, the local optimizers only converge to the nearest local minimum. The second Subfigure \ref{fig_cls_FS_progress} shows the sampling points (green) used in the bayesian optimization step of the FS optimizer and the selected final point for the local optimization step indicated by a red marker. The global optimizer finds one of the best possible minima.}
\label{fig_compare_landscapes}
\end{figure*}

On the other hand, local optimization focuses on finding the best solution within a specific region of the search space, usually centered around an initial guess or starting point. Unlike global optimization, local methods analyse the local behavior of the cost function near the starting point. While local optimization methods are generally faster and more efficient than global optimizers, as they do not explore the entire search space, they can get trapped in local minima. Additionally, their effectiveness is greatly dependent on their initialization.

Often there is an interplay between both types, as global optimizers can use local ones to improve the potential solutions found. There is no clear cut that defines all algorithms as part of one of the two groups. Methods like the Univariate Marginal Distribution Algorithm (UMDA) \cite{Fieldsend.07092022} (considered as a local optimizer in this work), could be argued to also sample a part of the parameter space, which is a feature of global methods. A brief overview over all used local and global optimizers and their categorization can be found in the Appendix \ref{appen_ah_loc} and \ref{appen_ah_glob}.

\begin{figure}[ht!]
     \centering
     \begin{subfigure}[b]{0.45\textwidth}
         \centering
         \includegraphics[width=\textwidth, trim={0cm 0cm 0cm 0cm},clip]{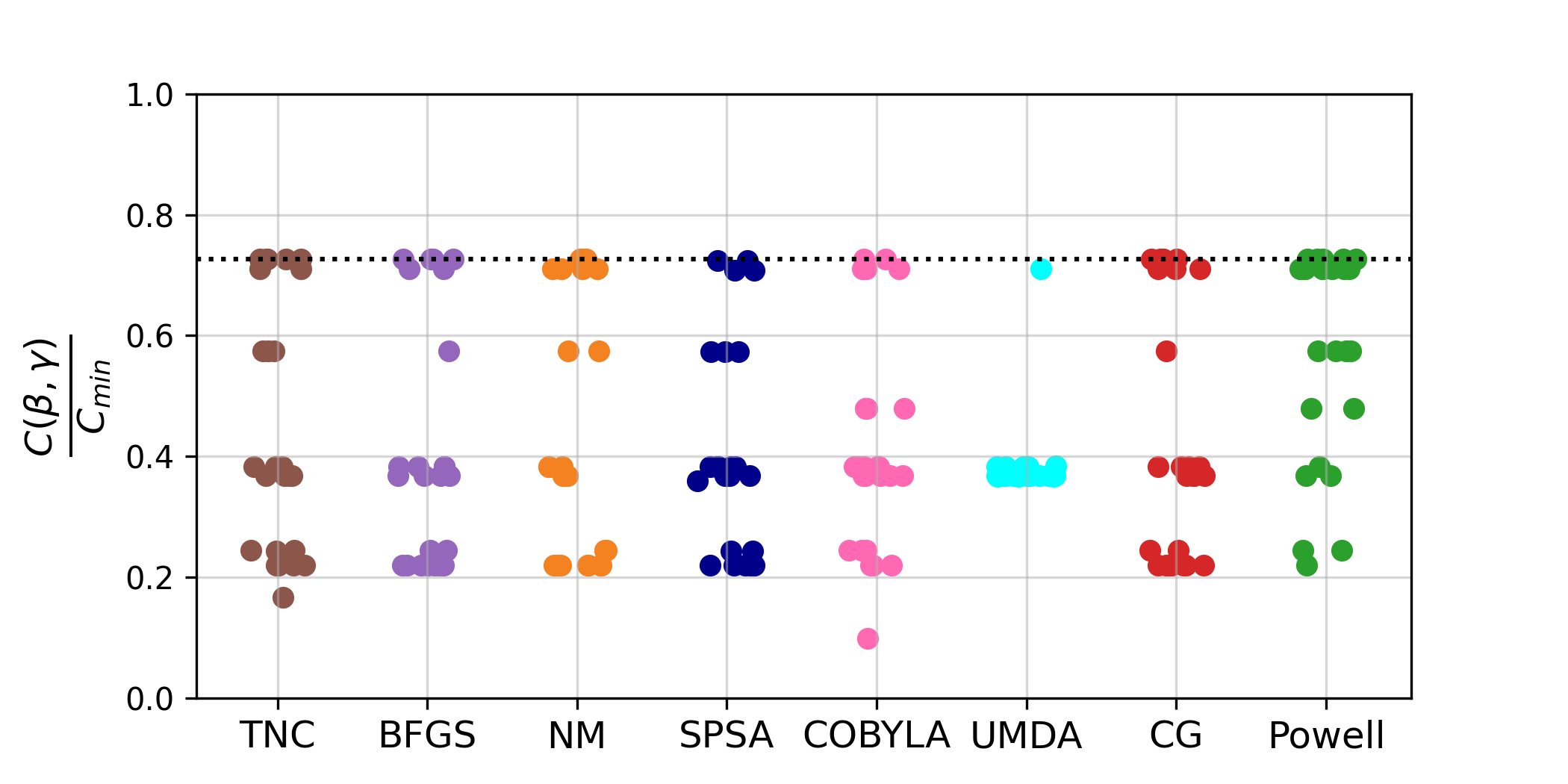}
     \end{subfigure}

     \begin{subfigure}[b]{0.45\textwidth}
         \centering
         \includegraphics[width=\textwidth, trim={0cm 0cm 0cm 0cm},clip]{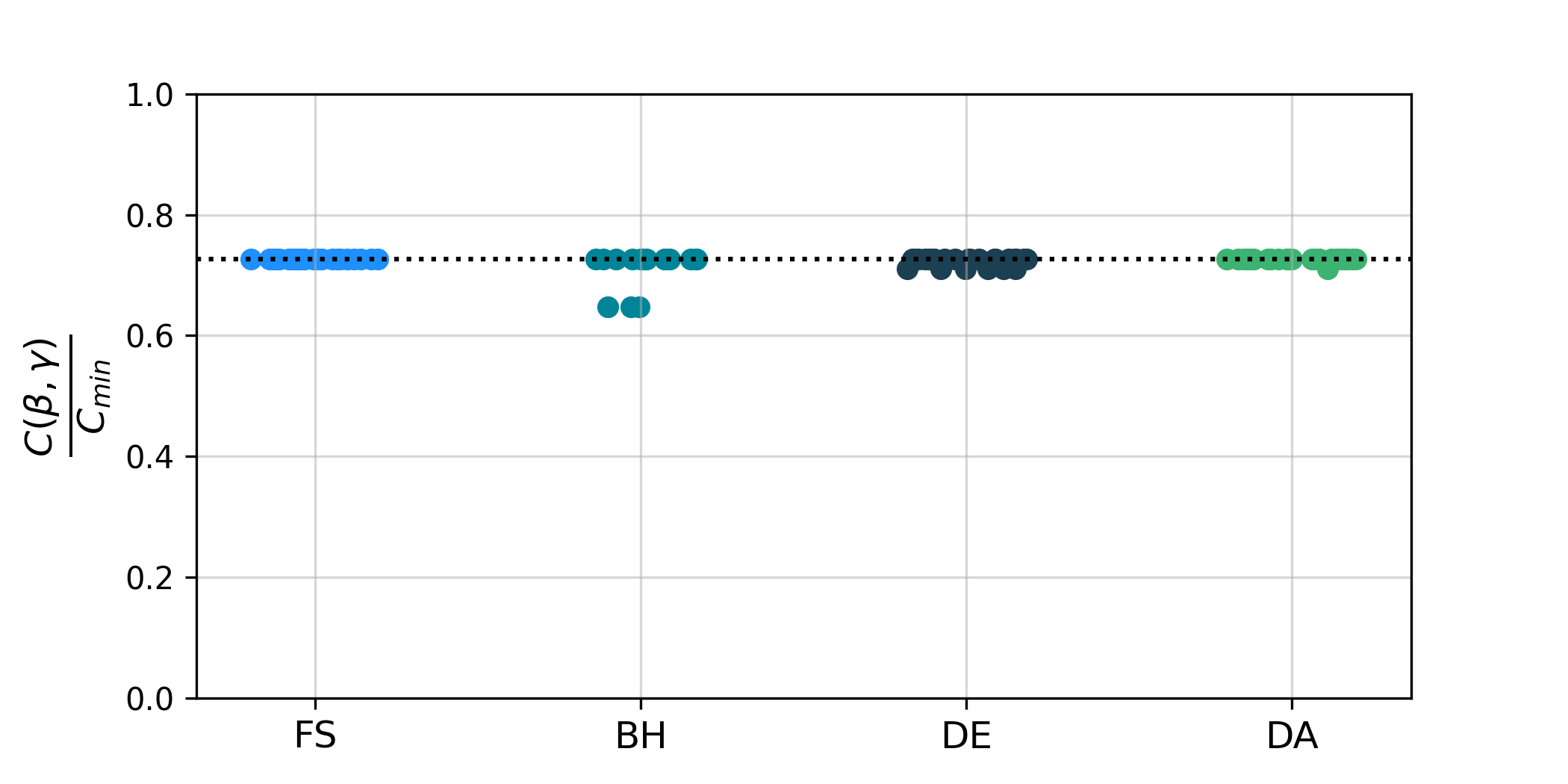}
     \end{subfigure}

        \caption{Comparison of local (upper) and global (lower) optimizers on the baseline cost landscape (four unit UC problem, $L=300$, $s=0.0002$ and $P=0.2$): The plots depict the results of 25 randomly initilizated runs of eight local and four global optimizers (see Appendix \ref{append_algs_heus}). The dotted line denotes the value of the best possible minimum in the defined search space. The local optimizers get stuck in various local minima depending on their initializations, while global optimizers mostly find the optimal solution.}\label{fig_comp_opt_res}
\end{figure}

In the following we illustrate the drawbacks of local optimization for QAOA and motivate the use of global optimizers with a simple example. In Figure \ref{fig_compare_landscapes} the cost landscape of an UC problem instance (previously depicted in Figure \ref{fig_pd_base}) can be seen. This cost landscape has been designed and optimized by the methods proposed by \cite{Brandhofer.2023, Shaydulin.2023}, as shown in Section \ref{chap_cl_cost_topology}, therefore representing a \textit{best case}. A local optimization algorithm (represented here by the Nelder-Mead (NM) optimizer) is initialized at four different initial points and run until convergence. One can clearly see, that the local optimizer gets trapped in local minima, even for such simple problem instance. Their effectiveness, compared to global optimizers, is therefore greatly dependent on their initialization. In contrast to this, Figure \ref{fig_cls_FS_progress} shows the behavior of the global optimization algorithm Fast-Slow (FS) \cite{Rad.04.03.2022}. This algorithm first samples globally and approximates the cost landscape by bayesian optimization. In a second step, a local optimizer is initialized at the most promising point. The yellow dots in Figure \ref{fig_cls_FS_progress} show the sampled points used for the global search, with the red dot denoting the most promising point for a local optimizer to start at. The global optimization phase ensures that the optimization algorithm is not trapped in local minima due to its (disadvantageous) initialization. 

To illustrate this point further, different optimizers are used to find the minimum in the cost landscape of Figure \ref{fig_comp_opt_res} with similar results. All local optimizers get trapped in various local minima, depending on their point of initialization. On the other hand, all global optimizers (almost always) find the global minima in the predefined parameter space. Note that in this scenario a value of 1 is not achievable, as the depth of QAOA as well as the choice of the parameter subspace does not allow such a solution. The improvement comes at the cost of an increasing number of function evaluations and time the optimization process takes. While local optimizers can achieve convergence within a few hundred function evaluations for simple problem instances, global optimizers might require several thousand evaluations to converge, if provided with the opportunity to do so. As this would take prohibitively long in simulations and especially in experiments on real hardware, this will be an important factor when comparing the different optimizers. 

%% file: experiments.tex
\section{Experiments}\label{chap_exp}

In order to demonstrate the advantage of global optimization, experiments on three industrially motivated use cases are performed. Most experiments were conducted on different configurations of the UC problem, which has the benefit of being scaleable to arbitrary numbers of qubits.

\begin{figure*}[ht!]
    \centering 
\begin{subfigure}{0.3\textwidth}
  \includegraphics[width=\linewidth]{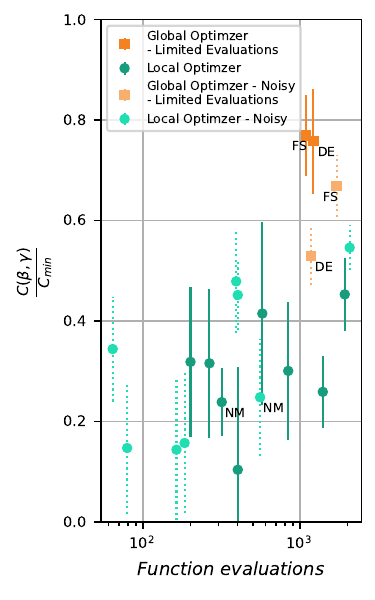}
  \caption{UC}
  \label{fig_cgl_UC}
\end{subfigure}\hfil 
\begin{subfigure}{0.3\textwidth}
  \includegraphics[width=\linewidth]{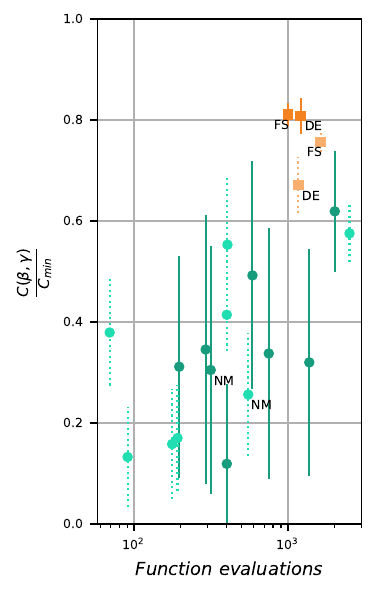}
  \caption{TSP}
  \label{fig_cgl_TSP}
\end{subfigure}\hfil 
\begin{subfigure}{0.3\textwidth}\hfil
  \includegraphics[width=\linewidth]{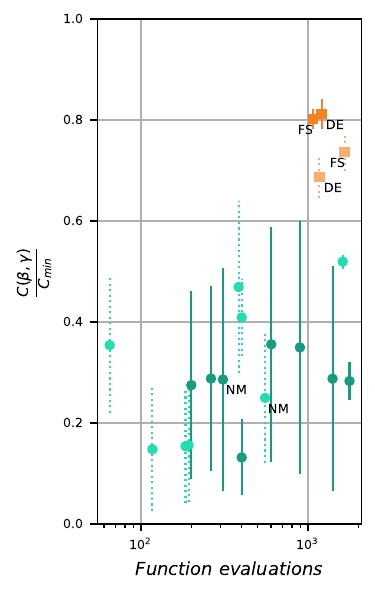}
  \caption{FL}
  \label{fig_cgl_FL}
\end{subfigure}\hfil 
\caption{Results of nine qubit use case instances of a three layer QAOA with adaptively chosen $s$ and, if needed, an iteratively chosen penalty factor. State vector and noisy simulations for all tested local optimizers and the most promising global optimizers with 10 random initializations each. NM is labeled as a reference to results in Table \ref{tab_hw_res}.}
\label{fig_comp_glob_loc}
\end{figure*}

\paragraph{Use Cases}
For all use cases, the cost landscapes of the problem instances are enhanced according to the principles of Section \ref{chap_cl_cost_topology} and kept constant for all experiments. Differences between samples are solely due to the optimizer used and their initialization points. The generalization capability of global optimizers is shown not only on varying problem instances, but across different use cases all together.

\paragraph{Optimizers}
Better solutions often come at the expense of higher number of function evaluations. These in turn influence the run time of an algorithm in simulations and on real hardware. Limiting the number of global optimization steps is therefore crucial. To balance this trade-off the global optimizers are partly set up such that results roughly require 1000 to 5000 function evaluations. Local optimizers are run for 200 optimization iterations each, with different numbers of function evaluations needed by each algorithm for one iteration. This restriction allows them to perform reasonably well. All optimizers described in Appendix \ref{append_algs_heus} are evaluated in simulations on the UC problem use case, and the best global optimizers are selected. These optimizers are evaluated on the TSP and FL use case and further investigated on quantum hardware.

\paragraph{Backend}
The simulations are carried out with an ideal statevector backend and a shot based noisy simulation backend. In the latter a custom noise model is used, which imitates the gate and readout errors described in qiskits \textit{FakeBoeblingenV2} backend, without copying its coupling map, therefore allowing for all-to-all connections in the circuit. This model results in lower error rates than the real counterpart, but allows for easier scaleable and faster experiments. To verify the findings from simulations, experiments are also run on real hardware. All experiments use layer numbers of $p=1$ and $p=3$.

\section{Results}\label{chap_results}
\subsection{Simulations}

\begin{table*}[ht!]
\fontsize{8pt}{8pt}\selectfont
\centering
\begin{tabular}{c | c | c | c  c | c c | c c | c c |}  
 & & & \multicolumn{2}{c|}{NM} & \multicolumn{2}{c|}{Powell}  & \multicolumn{2}{c|}{DE} & \multicolumn{2}{c|}{FS} \\ [1.2ex]
Q & p & back. & eval.  & norm. cost  & eval.  & norm. cost  & eval.  & norm. cost  & eval. & norm. cost   \\
\hline\hline
10 & 1 & s & 80$\pm$23 & 0.1$\pm$0.23 & 101$\pm$25 & 0.384$\pm$0.33 & 613$\pm$84 & 0.677$\pm$0.048 & 612$\pm$18 & 0.73$\pm$0.043 \\
10 & 1 & n & 525$\pm$8 & 0.261$\pm$0.116 & 105$\pm$54 & 0.421$\pm$0.145 & 1386$\pm$445 & 0.693$\pm$0.029 & 1489$\pm$224 & 0.715$\pm$0.017 \\
10 & 1 & e & 533 & 0.293 & - & - & 414 & 0.352 & 971 & 0.411 \\
\hline\hline
10 & 3 & s & 317$\pm$10	& 0.296$\pm$0.195 & 590$\pm$145 & 0.367$\pm$0.217 & 1254$\pm$85 & 0.767$\pm$0.048 & 1179$\pm$98 & 0.723$\pm$0.068\\
10 & 3 & n & 558$\pm$31	& 0.258$\pm$0.104 & 371$\pm$146 & 0.446$\pm$0.096 & 1201$\pm$27 & 0.493$\pm$0.032 & 1657$\pm$170 & 0.53$\pm$0.061\\
10 & 3 & e & 546 & 0.046 & - & - & 1137 & 0.11 & 974 & 0.092 \\ 
\hline\hline
14 & 3 & s & 322$\pm$14& 0.277$\pm$0.329 & 830$\pm$238 & 0.367$\pm$0.311 & 1230$\pm$32 & 0.877$\pm$0.002  & 1203$\pm$34 & 0.898$\pm$0.025 \\
14 & 3 & e & 549 & 0.091 & - & - & 1200 & 0.216 & 992 & 0.172 \\
\hline\hline
18 & 3 & s & 318$\pm$3 & 0.227$\pm$0.351 & 636$\pm$98 & 0.309$\pm$0.302 & 1192$\pm$147 & 0.922$\pm$0.022 & 830$\pm$104 & 0.905$\pm$0.001 \\
18 & 3 & e &582 & 0.132 & - & - & 1270 & 0.142 & 990 & 0.142 \\
\hline\hline

\end{tabular}
\caption{Results of hardware experiments on \textit{IBMQ Ehningen} (backend=e) and comparison with state-vector (backend=s) and noisy (backend=n) simulation on various UC problem instances with 10 to 18 qubits (Q). Two local optimizers (NM and Powell) are benchmarked against the \textit{restricted} global optimizers FS and DE in terms of function evaluations (eval.) and normalized cost (norm. cost).}
\label{tab_hw_res}
\end{table*}

All optimizers are tested on 8, 10 and 14 qubit UC problem instances  with 25 random initializations. The results for state vector simulations are depicted in Figure \ref{fig_results_opt_run} in  Appendix \ref{appendix_figure}. The global optimizers are evaluated with and without a limited number of function evaluations. In the following, just the abbreviations of the algorithms will be used (see Appendix \ref{append_algs_heus} for details). On all problem instances the \textit{unrestricted} global optimizers outperform the local optimizers, while requiring one to three orders of magnitude more function evaluations. The performance of the global optimizers SHGO and BH is greatly decreased, as the number of allowed function evaluations is reduced to 1000 to 5000. This is also the case for DE and DA, albeit this trend is attenuated especially for DE. The decrease in function evaluations leads especially for DA to a higher variance in performance across runs. Therefore the global optimizers FS and DE show the most promising behavior: high performance with low variance and relatively low number of function evaluations. These global optimizers are benchmarked in a  \textit{restricted} setting against all local optimizers on all three use cases with state vector as well as noisy simulations. In Figure \ref{fig_comp_glob_loc} the results for nine qubit instances of each use case are depicted. Here a clear trend can be observed: The \textit{restricted} global optimizers improve the performance of QAOA by a factor of two while increasing the number of function evaluations by roughly the same factor. At the same time the variance of performance is strongly reduced.

\subsection{Hardware experiments}

To verify the results of our simulations, we evaluated FS and DE as well as the local optimizer NM on the 27 qubits \textit{IBMQ Ehningen} quantum computer. The results in Table \ref{tab_hw_res} show that for all tested problem instances the global optimizers FS and DE outperform all tested local optimizers, while generally requiring more function evaluations. Among the different configurations, a deteriorating effect on the solution quality can be observed that is caused by hardware errors: The errors of deeper circuits even prevail the benefits of higher layer numbers, such that QAOA with $p=1$ has higher values of $\frac{C(\beta,\gamma)}{C_{min}}$ than its three layer counterpart. These findings are in line with the results in \cite{Harrigan.2021}, where - depending on the problem - better performance with increasing layer number is not to be expected. The limited number of samples available for higher numbers of qubits makes it hard to estimate the scalability of this approach, so it remains to be shown, that the advantage holds for larger systems, even though the trend seems promising.

%% file: conclusion.tex
\section{Conclusion}\label{chap_conclusion}

In this work, we investigated the performance of local optimizers on well designed cost landscapes of QAOA. We demonstrated that even for simple problem instances, local optimizers fail to find optimal solutions. Global optimizers on the other hand outperform local optimizers not only on such simple problem instances, but on a variety of use cases at the cost of higher numbers of function evaluations. These results hold for state vector simulations, noisy simulations as well as for experiments on real hardware. In order to overcome the caveat of high numbers of function evaluations required, we propose to restrict the global optimizers in terms of function evaluations. While these restrictions lead to a great decrease in number of function evaluations, the overall solution quality is only slightly decreased. Global optimizers like FS and DE can greatly improve QAOA's performance, while increasing the number of function evaluations only by a factor of two compared to local optimizers like Powell or NM. 

In current research, local optimization has been the standard approach to optimize the classical parameters of QAOA. Our work showed that restricted global optimizers are better suited and should be applied more widely. Especially combined with further enhancements in problem encodings, problem scalings and mixer designs, global optimizers can help to pave the way for the application of QAOA to real world problems.

%% file: appen_opt.tex
\section{Problem Formulations}\label{appendix_problems}

\subsection{Unit Commitment Problem}

The \textit{Unit Commitment} (UC) problem  describes the allocation of power units to satisfy a given power demand $L$ for the lowest possible price. The formulation used here is modelled after \cite{Koretsky.17.10.202122.10.2021}: 

\begin{equation}\label{eq_Cost_UC}
H_{cost,UC}(\{x_i\})= \sum_{i=1}^{n_{units}}(A_i+B_ip_i+C_ip_i^2)x_i.
\end{equation}

$A_i$,$B_i$ and $C_i$ are fixed parameters, that encode the cost for producing an amount of power $p_i$, if the unit $x_i$ is turned on. In this work, $p_i$ is additionally fixed to constant integer values, which are not subject to the optimization process. $n_{units}$ denotes the over-all number of units of the problem and is equal to the number of required qubits. 

To Equation \ref{eq_Cost_UC} constraines are added, to ensure that the sum of the produced power is equal to the power demand $L$, which can be expressed as the penalty term

\begin{equation}\label{eq_Pen_UC}
H_{pen,UC,1}(\{x_{i}\})=(\sum_{i=1}^N p_i y_i -L)^2.
\end{equation} 

Equations \ref{eq_Cost_UC} and \ref{eq_Pen_UC} are used to construct the QUBO formulation using Equation \ref{eq_gen_QUBO}. 

\subsection{Travelling Salesperson Problem}\label{chap_cl_TSP}

The objective of the \textit{Travelling Salesperson} (TSP) problem is to find the shortest possible route for visiting a list of cities and returning to the initial point. The used implementation is taken from \cite{Palackal.19.04.2023} and is described there in greater detail. The binary variables  $x_{i,j}$ of the problem have two indicies, with $i$ being an identifier for the city and $j$ the point in time, when it is visited. Both are integers $n_{cit}$, with  $n_{cit}^2$ qubits needed to construct the QAOA circuit.
With this the problem can be described as

\begin{equation}\label{eq_Cost_TSP}
H_{cost,TSP}(\{x_{ij}\})= \sum_{i,i',j=1}^{n_{cit}}D_{ii'}x_{ij}x_{i'(j+1)}.
\end{equation}

$D$ is the adjacency matrix describing the distances between each city, with the element $D_{ii'}$ being the distance between the $i^{th}$ and  $i'^{th}$ city. So each movement between cities in consecutive time steps adds to the cost. The constraints are added with the penalty term 
\begin{equation}\label{eq_Pen_TSP}
H_{pen,TSP,1}(\{x_{ij}\})= \sum_{i=1}^{n_{cit}}(1- \sum_{j=1}^{n_{cit}}x_{ij})^2 +  \sum_{j=1}^{n_{cit}}(1- \sum_{i=1}^{n_{cit}}x_{ij})^2.
\end{equation}

Equation \ref{eq_Pen_TSP} ensures that each city is only visited once and only one city is visited each time step respectively. The results in \cite{Palackal.19.04.2023} show, that using the same penalty factor for both constraints works reasonably well, so this simplification will be used here as well. Combining Equations \ref{eq_Cost_TSP} and \ref{eq_Pen_TSP} in Equation \ref{eq_gen_QUBO} constructs the QUBO formulation.

\subsection{Factory Layout Problem}

The  \textit{Factory Layout} (FL) problem is based on \cite{Klar.2022}. A number of positions $n_{pos}$ on a factory floor plan are given and some number $n_{mach}$ of production units has to be placed as efficiently as possible to minimize the cost of transport between each machine. As in the previous case the distances between the positions can be summarized in an adjacency matrix $D$. The flow of material between each of the machines can be described by a transportation density matrix $T$, where each element $T_{ii'}$ describes the amount of material going from the $i^{th}$ and  $i'^{th}$ production cell. To construct the problem, the decision variables have two indices with $x_{i,j}$ being the machine $i$ placed on position $j$. Using these definitions the problem can be formulated as

\begin{equation}\label{eq_Cost_FL}
min H_{cost,FL}(\{x_{ij}\})= \sum_{i,i'=1}^{n_{mach}}\sum_{j,j'=1}^{n_{pos}}D_{jj'}T_{ii'}x_{ij}x_{i'j'}.
\end{equation}

The constrains in this problem are the limitation of each machine being used exactly once and each position being at most taken once. This results in two distinct penalty terms: Each machine being used is written as

\begin{equation}\label{eq_Pen1_FL}
 H_{pen,FL,1}(\{x_{ij}\})=\sum_{i=1}^{n_{mach}}(\sum_{j=1}^{n_{pos}}(x_{ij}-1)^2,
\end{equation}
while each position being taken once is encoded as
\begin{equation}\label{eq_Pen2_FL}
 H_{pen,FL,2}(\{x_{ij}\})=\sum_{j=1}^{n_{pos}}\sum_{i,i'=1}^{n_{mach}}x_{ij}x_{i'j}.
\end{equation}

\section{Algorithms and Heuristics}\label{append_algs_heus}

\subsection{Local Optimizers}\label{appen_ah_loc}

The used optimizers are summarized in short and if not noted otherwise, the standard Scipy implementation is used.

\begin{itemize}
\item \textbf{Nelder-Mead (NM)}\cite{.1965}: This algorithm is derivative-free and uses a simplex of $n+1$ points, with $n$ being the dimensionality of the problem. Based on the function values of the different vertices a new point is constructed, that should be closer to the next minimum. The used Scipy implementation uses the version described in \cite{Gao.2012}.
\item \textbf{Powell}\cite{.1964}: This algorithm is derivative-free and performs several line searches to find the next optimum. It is constructed as an iterative process of choosing lines to search, based on the currently known minimum and performing the actual searches. The Scipy implementation is a modified version, that has no closer description.
\item \textbf{Conjugate Gradient (CG)}\cite{Hestenes.1952}: This algorithm uses derivatives and is originally designed to find solutions for big linear systems of equations. Based on gradient information it chooses a subspace to optimize for the next step.
\item \textbf{Broyden-Fletcher-Goldfarb-Shannon (BFGS)}\cite{BROYDEN.1970}: This algorithm approximates the gradients iteratively. Based on those the direction of descent is chosen and a line search is performed.
\item \textbf{Truncated Newton (TNC)}\cite{Dembo.1983}: This algorithm uses a CG-Method to solve the Newton equations and update the parameters for the next iteration based on this. It is designed to handle non linear functions as well.
\item \textbf{Contrained Optimization BY Linear Approximation (COBYLA)}\cite{Powell.1994}: This algorithm is derivative-free and uses a linear approximation of the problem to find potential points. For this a simplex is constructed that forms the basis of the approximation. In each iteration, vertices can then be replaced based on the approximation to improve the simplex or to save a found vector.
\item \textbf{Simultaneous Perturbation Stochastic Approximation (SPSA)}\cite{Spall.2001}: This algorithm approximates the gradient to follow by using only two measurements, so it is independent of dimensionality. With this gradient the next step is calculated and the approximation is performed again. The implementation used can be found in the \textit{qiskit} package. 
\item \textbf{Continuous Univariate Marginal Distribution Algorithm (UMDA)}\cite{Fieldsend.07092022}: This algorithm populates an area with sampling points and uses the best as data in the approximation of the cost function with univariate normal distributions. From these combined distributions new points are drawn for the sampling of a new generation. It is part of the \textit{qiskit} package.
\end{itemize}

\subsection{Global Optimizers}\label{appen_ah_glob}
As with the local optimizers this work mostly utilizes optimizers from the Scipy library, with deviations noted in the following.
\begin{itemize}
\item \textbf{Basin-Hopping (BH)}\cite{Wales.1997}: This algorithm is initialized at some random point and tries to find the next local minimum with a local minimizer. After convergence the algorithm applies a random perturbation to the coordinates in order to find a different local minimum. Depending on the found function value the algorithm can chose to use this point as the basis for the next global step or reject the point and revert to a previous solution. The sequence of local optimizations and global steps is repeated until either the maximum number of steps is reached or some condition is met. The used implementation ignores if the local optimizer converged to a minimum or not, and decides on accepting a new point purely based on its function value.

\item \textbf{Differential Evolution (DE)}\cite{Storn.1997}: This algorithm uses a random population of points in the search space to find the minimum. From iteration to iteration it combines the vectors of members of the current population to form members of the next generation.

\item \textbf{Dual Annealing (DA)}\cite{Xiang.1997}: This algorithm is also known as \textit{Generalized Simulated Annealing}. It tests new random points, minimizes them with a local optimizer and rejects ones with higher function values than the current best point based on a random distribution. This distribution is annealed over time, so the condition of acceptance becomes stricter.

\item \textbf{Simplical Homology Global Optimization (SHGO)}\cite{Endres.2018}: This algorithm uses sampling points to build a complex for approximating the cost function and finding potential regions where the minima might lie. It optimizes all possible points with a local optimizer to find all local minima and returns them.

\item \textbf{Fast-Slow (FS)}\cite{Rad.04.03.2022}: This algorithm is a hybrid between a bayesian learning regime and a classical local optimizer. In a first  step, the energy landscape is approximated over a wide area via bayesian approximation. With this model the most promising point is identified through classical optimization. This point will be minimized on the actual function by a local optimizer in a second step. This approach is the only one that is not directly implemented in the Scipy library. Instead, the bayesian learning from the \textit{Scikit-learn} library is used for the first step, while the second step is done with the scipy optimizers. 
\end{itemize}

\subsection{Adaptive Choice of Penalty Factors $P_j$}\label{append_ah_P}
The goal of the algorithm mentioned in \cite{Brandhofer.2023} is setting $P_j$ just large enough, so the optimal solution becomes the global optimum of the function of Equation \ref{eq_gen_QUBO}. This does not necessarily mean, that all wrong (invalid) solutions have a higher value of the cost function than any of the valid solutions. 

For the construction of $P_j$ the cost structure of all combinations needs to be known beforehand. Out of this one can extract $H_{min,opt}$, the cost of the optimal solution and $\overline{H}_{val}$, the mean cost of the valid solutions. 
With this the algorithm can start at a $P_j=0$:
\begin{enumerate}
\item Determine the lowest cost of the wrong solutions $H_{min,wrong}$
\item Check if following equation is true:
\begin{equation}
H_{min,wrong}\geq\frac{1}{2}(H_{min,opt}+\overline{H}_{val})
\end{equation} 
If yes, the algorithm terminates.
\item If the equation of the previous steps is not true, add to $P$ the following:
\begin{equation}
\Delta P=\frac{\frac{1}{2}(H_{min,opt}+\overline{H}_{val})-H_{min,wrong}}{(\sum_{i=1}^N p_i y_i^* -L)^2}
\end{equation}
with $ y_i^*$ being the states of the wrong solutions with the lowest cost.
\end{enumerate}

This approach is called \textit{adaptive}, as it is adapting to the over all cost structure. It has the obvious drawback, that for this algorithm the cost structure has to be known beforehand and therefore can not be used in a real application. 

\section{Figures}\label{appendix_figure}

\begin{figure*}[h]
     \centering
    \begin{subfigure}{1\textwidth}
    \centering
     \includegraphics[width=0.6\textwidth, trim={0cm 0cm 0cm 0cm},clip]{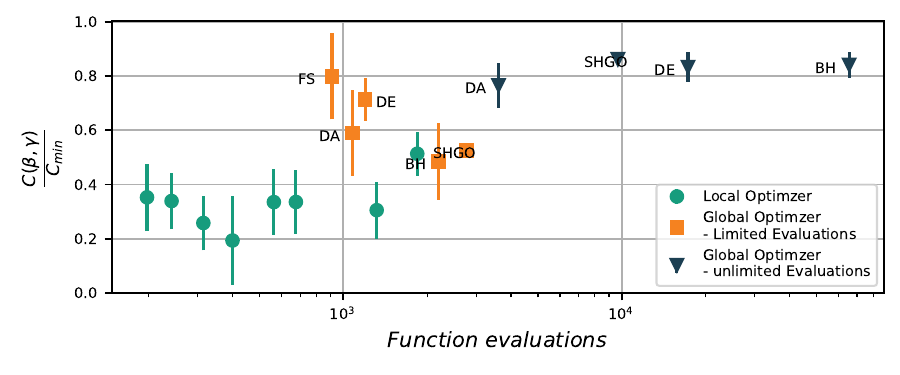}
	\end{subfigure}
	\begin{subfigure}{1\textwidth}
    \centering
     \includegraphics[width=0.6\textwidth, trim={0cm 0cm 0cm 0cm},clip]{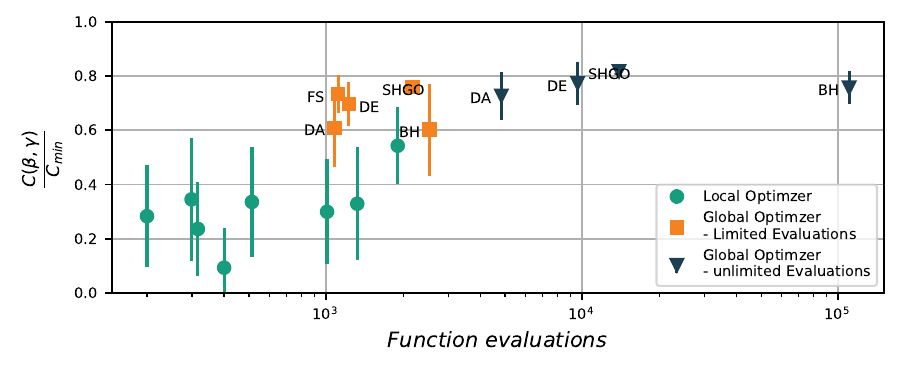}
	\end{subfigure}
    \begin{subfigure}{1\textwidth}
    \centering
     \includegraphics[width=0.6\textwidth, trim={0cm 0cm 0cm 0cm},clip]{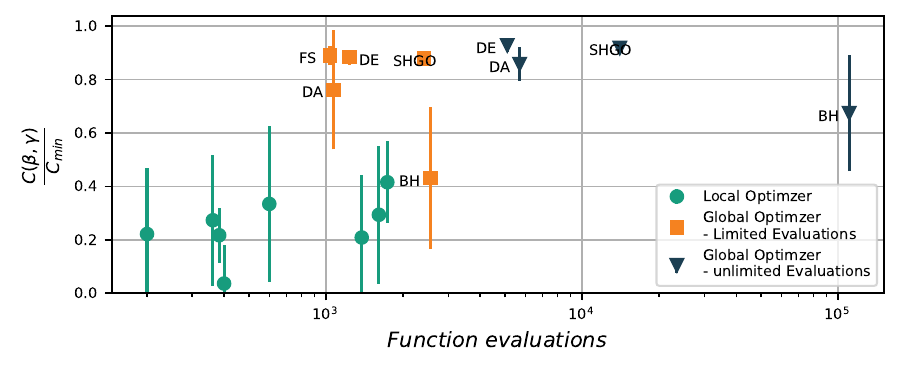}
	\end{subfigure}
	
     \caption{Results on three instances of the UC use case with 8 (upper), 10 (middle) and 14 (lower) qubits with a three layer QAOA. All optimizers are tested with 25 random initializations. Global optimizers are either run with restrictions of 1000 to 5000 function evaluations or unrestricted.}\label{fig_results_opt_run}
\end{figure*}

%% file: quantum-template.bbl
\begin{thebibliography}{10}

\bibitem{shor}
{Peter W. Shor}.
\newblock ``Polynomial-time algorithms for prime factorization and discrete
  logarithms on a quantum computer''.
\newblock SIAM Journal on Computing {\bf 26}, 1484–1509~(1997).
\newblock  url:~\url{https://epubs.siam.org/doi/10.1137/S0097539795293172}.

\bibitem{hhl}
{Aram Harrow, Avinatan Hassidim, Seth Lloyd}.
\newblock ``Quantum algorithm for solving linear systems of equations''.
\newblock Physical Review Letters.{\bf 15}~(2008).
\newblock  url:~\url{10.1103/PhysRevLett.103.150502}.

\bibitem{speed-up}
{G. Guerreschi, A. Matsuura}.
\newblock ``Qaoa for max-cut requires hundreds of qubits for quantum speed-
  up''.
\newblock Sci. Rep.{\bf 9}~(2019).
\newblock  url:~\url{https://doi.org/10.1038/s41598-019-43176-9}.

\bibitem{np}
{L. Bittel, M. Kliesch,}.
\newblock ``Training variational quantum algorithms is np-hard''.
\newblock Phys. Rev. Lett.{\bf 127}~(2021).
\newblock  url:~\url{https://doi.org/10.48550/arXiv.2101.07267}.

\bibitem{Ge.19.01.2022}
Xiaozhen Ge, Re-Bing Wu, and Herschel Rabitz.
\newblock ``The optimization landscape of hybrid quantum-classical algorithms:
  from quantum control to nisq applications''~(2022).

\bibitem{Brandhofer.2023}
Sebastian Brandhofer, Daniel Braun, Vanessa Dehn, Gerhard Hellstern, Matthias
  H{\"u}ls, Yanjun Ji, Ilia Polian, Amandeep~Singh Bhatia, and Thomas Wellens.
\newblock ``Benchmarking the performance of portfolio optimization with qaoa''.
\newblock \href{https://dx.doi.org/10.1007/s11128-022-03766-5}{Quantum
  Information Processing {\bf 22}, 34}~(2023).

\bibitem{Zhou.2020}
Leo Zhou, Sheng-Tao Wang, Soonwon Choi, Hannes Pichler, and Mikhail~D. Lukin.
\newblock ``Quantum approximate optimization algorithm: Performance, mechanism,
  and implementation on near-term devices''.
\newblock \href{https://dx.doi.org/10.1103/PhysRevX.10.021067}{Physical Review
  X {\bf 10}, 46}~(2020).

\bibitem{Nakanishi.2020}
Ken~M. Nakanishi, Keisuke Fujii, and Synge Todo.
\newblock ``Sequential minimal optimization for quantum-classical hybrid
  algorithms''.
\newblock \href{https://dx.doi.org/10.1103/PhysRevResearch.2.043158}{Physical
  Review Research {\bf 2}, 83}~(2020).

\bibitem{Sack.2021}
Stefan~H. Sack and Maksym Serbyn.
\newblock ``Quantum annealing initialization of the quantum approximate
  optimization algorithm''.
\newblock \href{https://dx.doi.org/10.22331/q-2021-07-01-491}{Quantum {\bf 5},
  491}~(2021).

\bibitem{Streif.23.08.2019}
Michael Streif and Martin Leib.
\newblock ``Training the quantum approximate optimization algorithm without
  access to a quantum processing unit''~(23.0).

\bibitem{Koretsky.17.10.202122.10.2021}
Samantha Koretsky, Pranav Gokhale, Jonathan~M. Baker, Joshua Viszlai, Honghao
  Zheng, Niroj Gurung, Ryan Burg, Esa~Aleksi Paaso, Amin Khodaei, Rozhin
  Eskandarpour, and Frederic~T. Chong.
\newblock ``Adapting quantum approximation optimization algorithm (qaoa) for
  unit commitment''.
\newblock In 2021 IEEE International Conference on Quantum Computing and
  Engineering (QCE).
\newblock \href{https://dx.doi.org/10.1109/QCE52317.2021.00035}{Pages
  181--187}.
\newblock IEEE~(17.10.2021 - 22.10.2021).

\bibitem{Palackal.19.04.2023}
Lilly Palackal, Benedikt Poggel, Matthias Wulff, Hans Ehm, Jeanette~Miriam
  Lorenz, and Christian Mendl.
\newblock ``Quantum-assisted solution paths for the capacitated vehicle routing
  problem''~(2023).

\bibitem{Klar.2022}
Matthias Klar, Philipp Schworm, Xiangqian Wu, Moritz Glatt, and Jan~C. Aurich.
\newblock ``Quantum annealing based factory layout planning''.
\newblock \href{https://dx.doi.org/10.1016/j.mfglet.2022.03.003}{Manufacturing
  Letters {\bf 32}, 59--62}~(2022).

\bibitem{global_op1}
C.A. Foudas and P.M. Pardalos, editors.
\newblock ``State of the art in global optimization: Computational methods and
  applications''.
\newblock \href{https://dx.doi.org/10.1007/978-1-4613-3437-8}{{Kluwer Academic
  Publishers}}. Dordrecht~(2013).

\bibitem{global_op2}
{Mohammad Nabi Omidvar, Xiaodong Li, Xin Yao}.
\newblock ``A review of population-based metaheuristics for large-scale
  black-box global optimization—part i''.
\newblock IEEE Transactions on Evolutionary Computation {\bf 26},
  802–822~(2021).
\newblock  url:~\url{https://doi.org/10.1109/TEVC.2021.3130838}.

\bibitem{global_op3}
{Bernd Hartke}.
\newblock ``Global optimization''.
\newblock Wiley Interdisciplinary Reviews: Computational Molecular Science {\bf
  26}, 879--887~(2011).
\newblock  url:~\url{https://doi.org/10.1002/wcms.70}.

\bibitem{Shaydulin.2023}
Ruslan Shaydulin, Phillip~C. Lotshaw, Jeffrey Larson, James Ostrowski, and
  Travis~S. Humble.
\newblock ``Parameter transfer for quantum approximate optimization of weighted
  maxcut''.
\newblock \href{https://dx.doi.org/10.1145/3584706}{ACM Transactions on Quantum
  Computing {\bf 103}, 1}~(2023).

\bibitem{Harrigan.2021}
Matthew~P. Harrigan, Kevin~J. Sung, Matthew Neeley, Kevin~J. Satzinger, Frank
  Arute, Kunal Arya, Juan Atalaya, Joseph~C. Bardin, Rami Barends, Sergio
  Boixo, Michael Broughton, Bob~B. Buckley, David~A. Buell, Brian Burkett,
  Nicholas Bushnell, Yu~Chen, Zijun Chen, Ben Chiaro, Roberto Collins, William
  Courtney, Sean Demura, Andrew Dunsworth, Daniel Eppens, Austin Fowler, Brooks
  Foxen, Craig Gidney, Marissa Giustina, Rob Graff, Steve Habegger, Alan Ho,
  Sabrina Hong, Trent Huang, L.~B. Ioffe, Sergei~V. Isakov, Evan Jeffrey, Zhang
  Jiang, Cody Jones, Dvir Kafri, Kostyantyn Kechedzhi, Julian Kelly, Seon Kim,
  Paul~V. Klimov, Alexander~N. Korotkov, Fedor Kostritsa, David Landhuis, Pavel
  Laptev, Mike Lindmark, Martin Leib, Orion Martin, John~M. Martinis, Jarrod~R.
  McClean, Matt McEwen, Anthony Megrant, Xiao Mi, Masoud Mohseni, Wojciech
  Mruczkiewicz, Josh Mutus, Ofer Naaman, Charles Neill, Florian Neukart,
  Murphy~Yuezhen Niu, Thomas~E. O'Brien, Bryan O'Gorman, Eric Ostby, Andre
  Petukhov, Harald Putterman, Chris Quintana, Pedram Roushan, Nicholas~C.
  Rubin, Daniel Sank, Andrea Skolik, Vadim Smelyanskiy, Doug Strain, Michael
  Streif, Marco Szalay, Amit Vainsencher, Theodore White, Z.~Jamie Yao, Ping
  Yeh, Adam Zalcman, Leo Zhou, Hartmut Neven, Dave Bacon, Erik Lucero, Edward
  Farhi, and Ryan Babbush.
\newblock ``Quantum approximate optimization of non-planar graph problems on a
  planar superconducting processor''.
\newblock \href{https://dx.doi.org/10.1038/s41567-020-01105-y}{Nature Physics
  {\bf 17}, 332--336}~(2021).

\bibitem{Khairy.2020}
Sami Khairy, Ruslan Shaydulin, Lukasz Cincio, Yuri Alexeev, and Prasanna
  Balaprakash.
\newblock ``Learning to optimize variational quantum circuits to solve
  combinatorial problems''.
\newblock \href{https://dx.doi.org/10.1609/aaai.v34i03.5616}{Proceedings of the
  AAAI Conference on Artificial Intelligence {\bf 34}, 2367--2375}~(2020).

\bibitem{genetic}
{Acampora, Giovanni, Angela Chiatto, and Autilia Vitiello.}
\newblock ``Genetic algorithms as classical optimizer for the quantum
  approximate optimization algorithm.''.
\newblock Applied Soft Computing{\bf 142}~({2023}).
\newblock
  url:~\url{https://www.sciencedirect.com/science/article/abs/pii/S1568494623003149}.

\bibitem{Rad.04.03.2022}
Ali Rad, Alireza Seif, and Norbert~M. Linke.
\newblock ``Surviving the barren plateau in variational quantum circuits with
  bayesian learning initialization''~(04.0).

\bibitem{warm-start}
{Daniel J. Egger, Jakub Marecek, and Stefan Woerner.}
\newblock ``Warm-starting quantum optimization''.
\newblock Quantum{\bf 5}~({2021}).
\newblock  url:~\url{https://doi.org/10.22331/q-2021-06-17-479}.

\bibitem{Glover.13.11.2018}
Fred Glover, Gary Kochenberger, and {Du Yu}.
\newblock ``A tutorial on formulating and using qubo models''~(13.1).

\bibitem{Hodson.13.11.2019}
Mark Hodson, Brendan Ruck, Hugh Ong, David Garvin, and Stefan Dulman.
\newblock ``Portfolio rebalancing experiments using the quantum alternating
  operator ansatz''~(13.1).

\bibitem{Li.28.12.2017}
Hao Li, Zheng Xu, Gavin Taylor, Christoph Studer, and Tom Goldstein.
\newblock ``Visualizing the loss landscape of neural nets''~(28.1).

\bibitem{Stechy.23.05.2023}
Micha{\l} St{\k{e}}ch{\l}y, Lanruo Gao, Boniface Yogendran, Enrico Fontana, and
  Manuel Rudolph.
\newblock ``Connecting the hamiltonian structure to the qaoa energy and fourier
  landscape structure''~(23.0).

\bibitem{Jing.2023}
Hang Jing, Ye~Wang, and Yan Li.
\newblock ``Data-driven quantum approximate optimization algorithm for power
  systems''.
\newblock \href{https://dx.doi.org/10.1038/s44172-023-00061-8}{Communications
  Engineering {\bf 2}, 1704}~(2023).

\bibitem{Baker.14.02.2022}
Jack~S. Baker and Santosh~Kumar Radha.
\newblock ``Wasserstein solution quality and the quantum approximate
  optimization algorithm: A portfolio optimization case study''~(2022).

\bibitem{Fieldsend.07092022}
Jonathan~E. Fieldsend and Markus Wagner, editors.
\newblock \href{https://dx.doi.org/10.1145/3520304}{``Proceedings of the
  genetic and evolutionary computation conference companion''}.
\newblock New York, NY, USA~(07092022).
\newblock ACM.

\bibitem{.1965}
{Nelder John A., Roger Mead}.
\newblock ``A simplex method for function minimization''.
\newblock The computer journal {\bf 7}, 308--313~(1965).
\newblock  url:~\url{https://doi.org/10.1093/comjnl/8.1.27}.

\bibitem{Gao.2012}
Fuchang Gao and Lixing Han.
\newblock ``Implementing the nelder-mead simplex algorithm with adaptive
  parameters''.
\newblock \href{https://dx.doi.org/10.1007/s10589-010-9329-3}{Computational
  Optimization and Applications {\bf 51}, 259--277}~(2012).

\bibitem{.1964}
{Powell Michael JD}.
\newblock ``An efficient method for finding the minimum of a function of
  several variables without calculating derivatives''.
\newblock The computer journal {\bf 7}, 155--162~(1964).
\newblock  url:~\url{https://doi.org/10.1093/comjnl/7.2.155}.

\bibitem{Hestenes.1952}
M.~R. Hestenes and E.~Stiefel.
\newblock ``Methods of conjugate gradients for solving linear systems''.
\newblock \href{https://dx.doi.org/10.6028/jres.049.044}{Journal of Research of
  the National Bureau of Standards {\bf 49}, 409}~(1952).

\bibitem{BROYDEN.1970}
C.~G. BROYDEN.
\newblock ``The convergence of a class of double-rank minimization algorithms
  1. general considerations''.
\newblock \href{https://dx.doi.org/10.1093/imamat/6.1.76}{IMA Journal of
  Applied Mathematics {\bf 6}, 76--90}~(1970).

\bibitem{Dembo.1983}
Ron~S. Dembo and Trond Steihaug.
\newblock ``Truncated-newton algorithms for large-scale unconstrained
  optimization''.
\newblock \href{https://dx.doi.org/10.1007/BF02592055}{Mathematical Programming
  {\bf 26}, 190--212}~(1983).

\bibitem{Powell.1994}
M.~J.~D. Powell.
\newblock ``A direct search optimization method that models the objective and
  constraint functions by linear interpolation''.
\newblock In Susana Gomez and Jean-Pierre Hennart, editors, Advances in
  Optimization and Numerical Analysis.
\newblock \href{https://dx.doi.org/10.1007/978-94-015-8330-5{\textunderscore
  }4}{Volume~7, pages 51--67}.
\newblock {Springer Netherlands}, Dordrecht~(1994).

\bibitem{Spall.2001}
{Spall James C.}
\newblock ``An overview of the simultaneous perturbation method for efficient
  optimization''~(1998).

\bibitem{Wales.1997}
David~J. Wales and Jonathan P.~K. Doye.
\newblock ``Global optimization by basin-hopping and the lowest energy
  structures of lennard-jones clusters containing up to 110 atoms''.
\newblock \href{https://dx.doi.org/10.1021/jp970984n}{The Journal of Physical
  Chemistry A {\bf 101}, 5111--5116}~(1997).

\bibitem{Storn.1997}
Rainer Storn and Kenneth Price.
\newblock ``Differential evolution: a simple and efficient heuristic for global
  optimization over continuous spaces''.
\newblock \href{https://dx.doi.org/10.1023/A:1008202821328}{Journal of Global
  Optimization {\bf 11}, 341--359}~(1997).

\bibitem{Xiang.1997}
Y.~Xiang, D.Y Sun, W.~Fan, and X.G Gong.
\newblock ``Generalized simulated annealing algorithm and its application to
  the thomson model''.
\newblock \href{https://dx.doi.org/10.1016/S0375-9601(97)00474-X}{Physics
  Letters A {\bf 233}, 216--220}~(1997).

\bibitem{Endres.2018}
Stefan~C. Endres, Carl Sandrock, and Walter~W. Focke.
\newblock ``A simplicial homology algorithm for lipschitz optimisation''.
\newblock \href{https://dx.doi.org/10.1007/s10898-018-0645-y}{Journal of Global
  Optimization {\bf 72}, 181--217}~(2018).

\end{thebibliography}
